\documentstyle[12pt,aasms4]{article}

\def\be{\begin{equation}}
\def\ee{\end{equation}}
\def\bea{\begin{eqnarray}}
\def\eea{\end{eqnarray}}
\def\btable{\begin{table}}
\def\etable{\end{table}}
\def\btabular{\begin{tabular}}
\def\etabular{\end{tabular}}
\def\bdoc{\begin{document}}
\def\edoc{\end{document}}
\def\bfig{\begin{figure}}
\def\efig{\end{figure}}
\def\bsmall{\begin{small}}
\def\esmall{\end{small}}

\def\n{\noindent}
\def\bb{\left(}
\def\eb{\right)}
\def\bbs{\left[}
\def\ebs{\right]}
\def\bba{\left<}
\def\eba{\right>}
\def\bbsq{\left[}
\def\ebsq{\right]}
\def\bbcu{\left\{}
\def\ebcu{\right\}}
\def\bml{\left|}
\def\emr{\right|}

\def\etal{et al.\ }
\def\ie{i.\,e.\,, }
\def\eg{e.\,g.\,} 
\def\etc{etc.\ }
\def\viz{viz.\,, }

\def\gne #1#2{\ \vphantom{S}^{\raise-0.5pt\hbox{$\scriptstyle#1$}}_
{\raise0.5pt \hbox{$\scriptstyle#2$}}}
\def\gneq{\gne > \sim}
\def\lneq{\gne < \sim}

\def\mod #1{|#1|}
\def\vec#1{{\bf #1}}
\def\mvec#1{\mod{\vec{#1}}}
\def\mean #1{\bba #1 \eba}
\def\meanvsq#1{\mean{\vec{#1}^2}}
\def\dotp#1#2{\vec{#1}\cdot\vec{#2}}
\def\meandotp#1#2{\mean{\dotp{#1}{#2}}}

\def\reci #1{\frac {1} {#1}}
\def\ten#1#2{#1{\times10^{#2}}}
\def\onlyten#1{10^{#1}}
\def\deg{\unit{deg}}
\def\degsq{\unit{deg^2}}
\def\ster{\unit{steradian}}
\def\perc{\unit{percent}}
\def\percent{\unit{percent}}

\def\deri#1#2{\frac {d#1} {d#2}}
\def\deriemp#1{\frac {d} {d#1}}
\def\pderi#1#2{\frac {\partial#1} {\partial#2}}
\def\pderiemp#1{\frac {\partial} {\partial#1}}
\def\secderi#1#2{\frac {d^2#1} {d#2^2}}
\def\secpderi#1#2{\frac {\partial^2#1} {\partial#2^2}}

\def\unit #1{\,{\rm #1}}

\def\hz{\unit{Hz}}
\def\khz{\unit{kHz}}
\def\mhz{\unit{MHz}}
\def\ghz{\unit{GHz}}
\def\thz{\unit{Thz}}
\def\ev{\unit{eV}}
\def\kev{\unit{keV}}
\def\mev{\unit{MeV }}
\def\gev{\unit{GeV }}
\def\tev{\unit{TeV }}

\def\funit{\rm\,erg\,cm^{-2}\,sec^{-1}}
\def\pfunit{\unit{cm^{-2}\,sec^{-1}}}
\def\funith{\rm\,erg\,cm^{-2}\,sec^{-1}\,Hz^{-1}}
\def\funitb{\rm\,erg\,cm^{-2}\,sec^{-1}\ster^{-1}}
\def\punitb{\rm\,photons\,cm^{-2}\,sec^{-1}\ster^{-1}}
\def\lunit{\rm\,erg\,sec^{-1}}
\def\lunith{\rm\,erg\,sec^{-1}\,Hz^{-1}}
\def\whz{\unit{W\,Hz^{-1}}}
\def\whzs{\unit{W\,Hz^{-1}\,str^{-1}}}
\def\jy{\unit{Jy}}
\def\mjy{\unit{mJy}}
\def\mujy{\unit{\mu Jy}}
\def\jyu{\unit{W\,m^{-2}\,Hz}}
\def\erg{\unit{erg}}
\def\ergs{\unit{erg\,s^{-1}}}
\def\ergsh{\unit{erg\,s^{-1}\,Hz^{-1}}}
\def\mcrab{\unit{milliCrab}}
\def\ergsc{\unit{erg\,s^{-1}\,cm^{-3}}}
\def\ergcmt{\unit{erg\,cm^{-3}}}
 
\def\mas{\unit{mas}}
\def\muas{\unit{{\mu}}as}
\def\masyr{\unit{mas\,yr^{-1}}}
\def\arcs{\unit{arcsec}}
\def\arcsec{\unit{arcsec}}
\def\arcsecsqi{\unit{arcsec^{-2}}}
\def\arcm{\unit{arcmin}}
\def\degsq{\unit{deg^2}}
\def\degsqi{\unit{deg^{-2}}}

\def\cm{\unit{cm}}
\def\nm{\unit{nm}}
\def\mm{\unit{mm}}
\def\cmsq{\unit{cm^2}}
\def\cmsqi{\unit{cm^{-2}}}
\def\cmcui{\unit{cm^{-3}}}
\def\mtr{\unit{m}}
\def\km{\unit{km}}
\def\pc{\unit{pc}}
\def\kpc{\unit{kpc}}
\def\mpc{\unit{Mpc}}
\def\gpc{\unit{Gpc}}
\def\sec{\unit{sec}}
\def\ksec{\unit{ksec}}
\def\msec{\unit{msec}}
\def\musec{\unit{\mu\,sec}}
\def\min{\unit{min}}
\def\hr{\unit{hr}}
\def\day{\unit{day}}
\def\days{\unit{days}}
\def\week{\unit{week}}
\def\yr{\unit{yr}}
\def\myr{\unit{Myr}}
\def\gyr{\unit{Gyr}}
\def\micron{\unit{\mu m}}
\def\delto{\Delta t_{\rm o}}

\def\cms{\unit{cm\,s^{-1}}}
\def\kms{\unit{km\,s^{-1}}}

\def\melec{m_{\rm e}}
\def\mprot{m_{\rm p}}

\def\fine{\alpha_{\rm f}}

\def\perc{\unit{percent}}
\def\mag{\unit{magnitude}}
\def\mags{\unit{magnitudes}}
\def\angst{\unit{\AA}}

\def\kel{\unit{K}}

\def\lsol{L_{\odot}}
\def\msol{M_{\odot}}
\def\rsol{R_{\odot}}
\def\zsol{Z_{\odot}}

\def\meight{\bb\frac{M}{\onlyten{8}\msol}\eb}

\def\aop{\alpha_{\rm op}}
\def\ar{\alpha_{\rm R}}
\def\ax{\alpha_{\rm x}}
\def\ag{\alpha_{\gamma}}
\def\aire{\alpha_{2.2-25\micron}}
\def\asm{\alpha_{\rm sm}}
\def\onemal{1-\alpha}
\def\nuone{\nu_1}
\def\nutwo{\nu_2}
\def\nuz{\nu_0}
\def\nug{\nu_g}
\def\nuc{\nu_c}
\def\nua{\nu_a}
\def\nup{\nu_p}
\def\nux{\nu_{\rm x}}
\def\nuop{\nu_{\rm op}}
\def\nupz{\nu_{p0}}
\def\nuaop{\nu^{-\alpha_{op}}}
\def\nualpha{\nu^{-\alpha}}
\def\alnu{\alpha_{\nu}}
\def\alnua{\alpha_{\nu_a}}
\def\alnusc{\alnu^{\rm sc}}
\def\alsc{\alpha_{\rm sc}}
\def\jnu{j_{\nu}}
\def\jnua{j_{\nu_a}}
\def\taustar{\tau_{*}}
\def\tauab{\tau_{\rm ab}}
\def\tnu{\tau_{\nu}}
\def\fnu{F(\nu)}
\def\fnua{F(\nua)}
\def\fr{F_{\rm R}}
\def\fop{F_{\rm op}}
\def\fopl{F_{\rm op}^l}
\def\fx{F_{\rm x}}
\def\fxl{F_{\rm x}^{l}}
\def\frlim{F_{R,lim}}
\def\mv{m_V}
\def\mb{m_B}
\def\mr{m_R}
\def\Mv{M_V}
\def\Mb{M_B}
\def\Mr{M_R}
\def\inu{I(\nu)}
\def\Snu{S(\nu)}
\def\Snua{S(\nu_a)}
\def\Pnu{P(\nu)}
\def\lnu{L(\nu)}
\def\llnu{\log \lnu}
\def\lr{L_{\rm R}}
\def\llr{\log L_{\rm R}}
\def\lrlim{L_{R,lim}}
\def\lfive{L(5\ghz)}
\def\lop{L_{{\rm op}}}
\def\llop{\log L_{{\rm op}}}
\def\lx{L_{\rm x}}
\def\llx{\log L_{\rm x}}
\def\lxmin{L_{\rm x,min}}
\def\lxmax{L_{\rm x,max}}
\def\lmax{L_{|rm max}}
\def\lir{L_{\rm IR}}
\def\luv{L_{\rm UV}} 
\def\lrc{L_{\rm rc}}
\def\llrc{\log L_{\rm rc}}
\def\lrext{L_{\rm \rm r,ext}}
\def\llrext{\log L_{r,ext}}
\def\lre{L_{\rm re}}
\def\llre{\log L_{\rm re}}
\def\lxb{L_{\rm xb}}
\def\llxb{\log L_{\rm xb}}
\def\lxu{L_{\rm xu}}
\def\llxu{\log L_{\rm xu}}
\def\lre{L_{\rm re}}
\def\llre{\log L_{\rm re}}
\def\rt{R_{\rm t}}
\def\rx{R_{\rm x}}
\def\rxbu{R_{\rm xbu}}
\def\rtx{R_{\rm tx}}
\def\gxbp{g_{\rm x}(\beta, \psi)}
\def\grbp{g_{\rm r}(\beta, \psi)}
\def\aox{\alpha_{\rm ox}}
\def\aoxl{\alpha_{\rm ox}^l}
\def\aro{\alpha_{\rm ro}}
\def\arx{\alpha_{\rm rx}}
\def\aoxx{\alpha_{\rm oxx}}
\def\auv{\alpha_{\rm uv}}
\def\sigt{\sigma_{\rm T}}
\def\sigkn{\sigma_{\rm KN}}
\def\signu{\sigma_{\nu}}
\def\siggg{\sigma_{\gamma\gamma}}
\def\sigsc{\sigma_{\rm sc}}
\def\tausc{\tau_{\rm sc}}
\def\taut{\tau_{\rm T}}
\def\nsc{N_{\rm sc}}
\def\freenu{l_{\nu}}
\def\U #1{U_{\rm #1}}
\def\eps{\epsilon}
\def\epsone{\epsilon_1}
\def\epstwo{\epsilon_2}
\def\epsx{\epsilon_{\rm x}}
\def\epsxs{\eps_{\rm xs}}
\def\epss{\epsilon_{\rm s}}
\def\epsg{\epsilon_{\gamma}}
\def\epsnu{\epsilon_{\nu}}
\def\epsir{\eps_{\rm IR}}
\def\DOM{\Delta\Omega}

\def\vbyvm{\frac {V} {V_m}}
\def\vbyvmp{\frac {V'} {V_m'}}
\def\mvbyvm{\bba\vbyvm\eba}
\def\mvbyvmp{\bba\vbyvmp\eba}

\def\vbyvml{V/V_m}
\def\vbyvmpl{V'/V_m'}
\def\mvbyvml{\bba\vbyvml\eba}
\def\mvbyvmpl{\bba\vbyvmpl\eba}

\def\bperp{B_{\perp}}
\def\bpar{B_{\parallel}}
\def\bperppar{B_{\perp-\parallel}}
\def\dtvar{\Delta t_{var}}
\def\Tvar{T_{var}}
\def\dmv{dm_v}
\def\betaa{\beta_a}
\def\gammap{\gamma_{\rm p}}
\def\gammab{\gamma_{\rm b}}
\def\gammamax{\gamma_{\rm max}}
\def\scrid{{\cal D}}
\def\scril{{\cal L}}
\def\fbeam{F_b}
\def\fext{F_{ext}}
\def\bcospsi{\beta\cos\psi}
\def\bsinpsi{\beta\sin\psi}
\def\rnu{R_{\nu}}
\def\rnuone{R_{\nu_1}}
\def\rnutwo{R_{\nu_2}}

\def\nh{N_{\rm H}}
\def\eax{E^{-\ax}}
\def\ex{E_{\rm x}}
\def\ex{E_{\rm xs}}

\def\zmin{z_{\rm min}}
\def\zmax{z_{\rm max}}
\def\rmin{r_{\rm min}}
\def\rmax{r_{\rm max}}
\def\rsc{r_{\rm sc}}
\def\rscm{r_{\rm sc,m}}

\def\corrxz{r_{\rm x,z}}
\def\corrxop{r_{\rm x,op}}
\def\corropz{r_{\rm op,z}}
\def\corrxopz{r_{\rm x,op;z}}
\def\chisq{\chi^2}
\def\chisqr{\chi^{2}_{\nu}}

\def\philx{\Phi{(\lx)}}
\def\phixstar{\Phi_{X}^{*}}
\def\phixstaro{\Phi_{X1}^{*}}
\def\lxstar{\lx^{*}}
\def\lxfour{L_{{\rm x},44}}
\def\lxstarfour{L_{{\rm x},44}^*(0)}

\def\asca{{\em ASCA }}
\def\axaf{{\em AXAF }}
\def\ein{{\em EINSTEIN }}
\def\exosat{{\em EXOSAT }}
\def\ginga{{\em GINGA }}
\def\rosat{{\em ROSAT }}
\def\compton{{\em COMPTON }}
\def\granat{{\em GRANAT }}
\def\iras{{\em IRAS }}
\def\iue{{\em IUE }}

\def\seyone{Seyfert\,1 }
\def\seytwo{Seyfert\,2 }
\def\bllac{BL Lac }
\def\bllacs{BL Lacs }
\def\rbl{{\rm RBL }}
\def\xbl{{\rm XBL }}
\def\ipc{{\rm IPC }}
\def\pspc{{\rm PSPC }}

\def\cfour{{\rm C\,IV }}
\def\cthreesf{{\rm C\,III] }}
\def\fetwo{{\rm Fe\,II }}
\def\halpha{{\rm H\,\alpha }}
\def\hbeta{{\rm H\,\beta }}
\def\hgamma{{\rm H\,\gamma}}
\def\hi{{\rm H\,I }}
\def\lyal{{\rm Ly\,\alpha }}
\def\mgtwo{{\rm Mg\,II }}
\def\ntwo{{\rm [N\,II]}}
\def\nfive{{\rm N\,V }}
\def\otwo{{\rm [O\,II] }}
\def\othree{{\rm [O\,III] }}
\def\kalpha{K_{\alpha} }
\def\ykz{Y^{K}_{Z} }

\def\const{{\rm constant}}
\def\gray{\gamma-{\rm ray}}
\def\grays{\gamma-{\rm rays}}
\def\maxtau{\tausc(1+\tausc)}
\def\maxtnu{\tnu(1+\tnu)}
\def\dm{{\rm DM}}
\def\dmsinmb{\dm\sin\mod{b}}
\def\dmsinb{\dm\sin b}

\def\vbyc{\frac{v}{c}}
\def\vbycsq{\frac{v^2}{c^2}}
\def\omvbycsq{\sqrt{1-\vbycsq}}

\def\gmr{\frac{GM}{r}}
\def\gmcsr{\frac{GM}{c^2r}}
\def\gmcsrl{GM/c^2r}
\def\tgmcsr{\frac{2GM}{c^2r}}
\def\schwarz{r_{\rm S}}
\def\tschwarz{t_{\rm S}}
\def\rergo{r_{\rm E}}
\def\rplus{r_{+}}

\def\omk{\Omega_{\rm K}(r)}
\def\velr{v_{r}}
\def\velp{v_{\phi}}
\def\velz{v_{z}}
\def\sigrt{\Sigma(r,t)}
\def\mach{{\cal M}}
\def\scrir{{\cal R}}
\def\tin{t_{\rm in}}
\def\tl{t_{\rm L}}
\def\ldisk{L_{\rm disk}}
\def\temps{T_{\rm s}}
\def\tempc{T_{\rm c}}
\def\bnuts{B_{\nu}(\temps)}
\def\bnutsr{B_{\nu}(\temps(r))}
\def\rout{r_{\rm out}}
\def\tvis{t_{\rm vis}}
\def\pgas{p_{\rm g}}
\def\prad{p_{\rm r}}
\def\csou{c_{\rm s}}
\def\rstar{r_{*}}
\def\ledd{L_{\rm Edd}}
\def\mdedd{\dot{M}_{\rm Edd}}
\def\tedd{t_{\rm Edd}}
\def\tempedd{T_{\rm Edd}}
\def\bedd{B_{\rm Edd}}
\def\mdot{{\dot M}}
\def\geff{\vec{g}_{\rm eff}}
\def\phir{\Phi_{\rm rot}}
\def\phie{\Phi_{\rm eff}}
\def\zsr{z_{\rm s}(r)}
\def\udzero{u_0}
\def\udphi{u_{\phi}}
\def\uuzero{u^0}
\def\uuphi{u^{\phi}}

\def\CH{{\it CH}}

\def\hubble#1{H_0={#1}\unit{km\,sec^{-1}\,Mpc^{-1}}}
\def\hubbleunit{\unit{km\,sec^{-1}\,Mpc^{-1}}}
\def\hubhund{h_{100}}
\def\hubhundi{\hubhund^{-1}}
\def\hubhundis{\hubhund^{-2}}
\def\cuponh{\bb \frac {c} {H_0} \eb}

\def\qz{q_0}
\def\qzsq{\qz^2}
\def\fqz{\bbs \qz z + (q_0-1)(\sqrt{1+2q_0z}-1) \ebs}
\def\omb{\Omega_{{\rm b}}}
\def\omhi{\Omega_{{\rm H\,I}}}
\def\omlls{\Omega_{{\rm LLS}}}
\def\omlb{\Omega_{{\rm lb}}}
\def\omigm{\Omega_{{\rm IGM}}}
\def\omdlyal{\Omega_{{\rm DLy\,\alpha}}}
\def\zpr{z^{\prime}}
\def\zabs{z_{{\rm abs}}}
\def\zem{z_{{\rm em}}}
\def\dlum{D_{\rm L}}

\def\labchap #1{\label{chap:#1}}
\def\labequn #1{\label{eq:#1}}
\def\labfig #1{\label{fig:#1}}
\def\labsecn #1{\label{sec:#1}}
\def\labsubsecn #1{\label{subsecn:#1}}
\def\labsubsubsecn #1{\label{subsubsecn:#1}}
\def\labtablem #1{\label{tab:#1}}
\def\labapp #1{\label{app:#1}}

\def\chap #1{Chapter~\ref{chap:#1}}
\def\equn #1{Equation~\ref{eq:#1}}
\def\fig #1{Figure~\ref{fig:#1}}
\def\relation #1{relation~\ref{eq:#1}}
\def\secn #1{Section~\ref{sec:#1}}
\def\subsecn #1{Section~\ref{subsecn:#1}}
\def\subsubsecn #1{Section~\ref{subsubsecn:#1}}
\def\app#1{Appendix~\ref{app:#1}}
\def\tablem #1{Table~\ref{tab:#1}}

\def\dequn#1#2{Equations~{\ref{eq:#1}}~and~{\ref{eq:#2}}}
\def\tequn#1#2#3{Equations~{\ref{eq:#1}},~{\ref{eq:#2}} and ~{\ref{eq:#3}}}
\def\mequn#1#2{Equations~{\ref{eq:#1}}~to~{\ref{eq:#2}}}
\def\dfig#1#2{Figures~{\ref{fig:#1}}~and~{\ref{fig:#2}}}
\def\mfig#1#2{Figures~{\ref{fig:#1}}~to~{\ref{fig:#2}}}
\def\dtablem#1#2{Tables~{\ref{tab:#1}}~and~{\ref{tab:#2}}}
\def\dsubsecn#1#2{Subsections~{\ref{subsecn:#1}}~and~{\ref{subsecn:#2}}}

\def\aj#1{{\it Astron. J. }{\bf #1}}
\def\anp#1{{\it Ann. Phys. }{\bf #1}}
\def\apj#1{{\it Astrophys. J. }{\bf #1}}
\def\apjs#1{{\it Astrophys. J. Suppl. }{\bf #1}}
\def\apss#1{{\it Astrophys. Sp. Sci. }{\bf #1}}
\def\araa#1{{\it Ann. Rev. Astron. Astrophys. }{\bf #1}}
\def\asa#1{{\it Astro. Astrophys. }{\bf #1}}
\def\asas#1{{\it Astron. Astrophys. Suppl. }{\bf #1}}
\def\asp#1{{\it ASP Conf. Ser. }{\bf #1}}
\def\aujp#1{{\it Austr. J. Phys. }{\bf #1}}
\def\aar#1{{\it Astron. Astrophy. Rev }{\bf #1}}

\def\iau#1{{\it IAU Symp.}{\bf #1}}
\def\ijmp#1{{\it Int. J. Mod. Phy. D }{\bf #1}}
\def\jaa#1{{\it Journ. Astrophys. Astron.}}
\def\jetp#1{{\it Sov. Phys. JETP }{\bf #1}}

\def\mn#1{{\it Mon. Not. Roy. astr. Soc. }{\bf #1}}
\def\nat#1{{\it Nature }{\bf #1}}
\def\ns#1{{\it New Scientist }{\bf#1}}

\def\pasp#1{{\it Proc. Astr. Soc. Pacific }{\bf #1}}
\def\pasj#1{{\it Proc. Astron. Soc. Japan }{\bf #1}}
\def\pnas#1{{\it Proc. Nat. Acad. Sci. USA }{\bf #1}}
\def\plo#1{{\it Pub. Lick Obss. }{\bf #1}}
\def\pla#1{{\it Phys. Lett. A }{\bf #1}}
\def\phyr#1{{it Phys. Rep. }{\bf #1}} 
\def\prew#1{{\it Phys. Rev. }{\bf #1}}
\def\prewd#1{{\it Phys. Rev. D }{\bf #1}}
\def\prl#1{{\it Phys. Rev. Lett. }{\bf #1}}
\def\prsl#1{{\it  Proc. Roy. Soc. Lond. A }{\bf #1}}
\def\rmp#1{{\it Rev. Mod. Phys. }{\bf #1}}
\def\rep#1{{\it Rep. Prog. Phy. }{\bf #1}}
\def\sci#1{{\it Science }{\bf #1}}
\def\spsr#1{{\it Sp. Sci. Rev. }{\bf #1}}
\def\stsi#1{{\it Space Telescope Science Institute Preprint }{\bf #1}}
\def\zphy#1{{\it Zeits. Phys. }{\bf #1}}

\def\cfa#1{{\it Center for Astrophysics Preprint }{\bf #1}}
\def\asp{Astron. Soc. of the Pacific, San Francisco.}
\def\clar{Clarendon Press, Oxford.} 
\def\cup{Cambridge University Press, Cambridge.}
\def\free{W. H. Freeman, New York.}
\def\klu{Kluwer, Dordrecht.}
\def\nrao{NRAO, Greenbank.}
\def\plenumn{PLenum Press, New York.}
\def\prince{Princeton University Press, Princeston}
\def\ridel{Ridel, Dordrecht.}
\def\springerb{Springer-Verlag, Berlin.}
\def\springern{Springer-Verlag, New York.}
\def\wiley{Wiley, New York.}
\def\yale{Yale University Press, New Haven.}

\def\sqr#1#2{{\vcenter{\vbox{\hrule height.#2pt
  \hbox {\vrule width.#2pt height#1pt \kern#1pt
  \vrule width.#2pt}
  \hrule height.#2pt}}}}

\def\square{\mathchoice\sqr68\sqr68\sqr{2.1}3\sqr{1.5}3}

\def\CH{{\it CH }}

\begin{document}

\title{ Broadening of the Iron emission line in MCG-6-30-15 by Comptonization }

\author{\bf R. Misra}
\authoremail{rmisra@iucaa.ernet.in}
\author{\bf A. K. Kembhavi}
\affil{Inter-University Centre for Astronomy and Astrophysics, Pune, India}
\authoremail{akk@iucaa.ernet.in}

\begin{abstract}
We show that the Iron K emission line from MCG-6-30-15 could
be broadened due to Comptonization by a surrounding highly ionized
cloud with radius $\sim 10^{14}$ cms. We calculate
the temperature of the cloud to be $\sim0.21\kev$, provided a  reasonable 
estimate of the UV flux is made. The X-ray/$\gamma$-ray emission observed
from the source is compatible with this model. Such a cloud
should be highly ionized and strong absorption edges are
not expected from the source (Fabian \etal 1995).

For a $\onlyten{6}\msol$ black hole the size of the could corresponds to about
$300$ Schwarzschild radius. The intrinsic line could then be
emitted far from the black hole and gravitational red-shift and
Doppler effects would be negligible. If the black hole mass is much
larger than $\onlyten{6}\msol$, gravitational/Doppler red-shifts would
also contribute significantly to the broadening.

We argue that the broad red wing observed in the source
does not by itself imply emission from regions close ( $R < 5 r_s$) 
to the black hole. However, Comptonization cannot produce
a double peak. The presence of such a feature is a clear sign
of inner disk emission influenced by gravitational and Doppler effects,
perhaps broadened by the Comptonization. We note that simultaneous
broad band (2-100 keV) study of this source can also reveal (or rule out)
the presence of such a Comptonizing cloud.

\end{abstract}


\section{Introduction}

Recent ASCA observations have shown that the Iron K emission lines in
the spectra of many Seyfert 1 galaxies are broad (Tanaka \etal 1995;
Mushotzky \etal 1995). A long  
($\approx 4.5$ days) observation of the bright Seyfert 1 galaxy
MCG-6-30-15 showed that observed line width corresponds to a 
velocity $\sim\onlyten{10}\cms$ and is asymmetric with most of
the emission being red-shifted (Tanaka \etal 1995). Furthermore,
no strong absorption edges were observed from this source. 
 It was
realized that such  emission is expected when the line
is produced in the inner region $3 r_s < R < 10 r_s$ about
a black hole, where $r_s=2GM/c^2$ is the Schwarzschild radius. The line
shape is due to the combined effect of gravitational
and Doppler red-shifts ( Fabian \etal 1989 ). Spectral
fitting revealed that the line is variable and that during
one epoch ( the deep minimum phase ) it seems to be generated
close ($ \approx 2 r_s$) to a spinning (Kerr) black hole (Iwasawa \etal 1996).
If this interpretation is true this would be the first direct observation
of the strong gravitational effects expected close to a black hole. Moreover,
this result strongly constrains theoretical models. Since the cold
accretion disk extends all the way to the last stable orbit, models
having a hot inner disk (for e.g. Shapiro, Lightman \& Eardley 1976;
Misra \& Melia 1996; Narayan \& Yi 1994; Chakrabarti 1997) would no longer 
be valid for this source. At the same time
the X-ray source would need to be situated very close to the black hole 
($ \approx 2 r_s$). Models where the X-rays are produced in hot active
regions (or coronas) ( for e.g. Liang \& Price 1977; Haardt \& Maraschi
1993) above the cold disk would be more promising. However, at least for
MCG -6-30-15,
these active regions would now have to be concentrated near the last stable
orbit instead of an inner region ( $\approx 20 R_s$) as has been
hitherto assumed.

The importance of these results and their interpretation warranted the
examination of  alternative models which could explain the observations.
Fabian \etal (1995) considered several other kinds of mechanisms for the
broadening
and concluded that none of them was satisfactory, thus
strengthening the case for the inner disk emission model. One
of the possibilities considered by Fabian \etal (1995) was the broadening 
of a narrow intrinsic line by Comptonization 
(Czerny, Zbyszewska and Raine 1991).
They concluded that to explain the line profile the electron
temperature in the Comptonizing medium would have to be less
than $0.25\kev$ and have an optical depth $\tau = 5$. Moreover,
in order not to have large photo-electric edges, the region was required to
be highly ionized. Fabian \etal (1995) found this model unsatisfactory
because (a) for the medium to be highly ionized the region should be
less than $50\,r_s$ from a $\onlyten{7}\msol$ black hole, which would make 
gravitational
effects dominant,  (b) the X-ray emission would be altered by such a covering
medium,  and c) the Compton temperature of
this region is expected to be higher than $0.25\kev$, which is 
inconsistent with the
upper limit on the temperature obtained from the profile. 

In this paper we argue that the present X-ray/$\gamma$-ray observations
of this source are compatible with a covering Comptonizing cloud.
We calculate the self consistent temperature by balancing
input to the output power in this cloud and find that the temperature
is $\approx 0.2$ keV, provided a reasonable estimate of
the UV flux in this source is made. We show that the essential features
of the line, namely, a broad red wing and a sharp blue drop can also
be reproduced by a Comptonized line spectrum. Thus, it may be possible
that a Comptonizing cloud with radius $\approx 10^{14}$ cms surrounding
the black hole contributes significantly to the broadening of the Iron
line.  In this case, the innermost region of an accretion disk ($ r < 10 r_s$)
could still be hot, with the reflection arising from the inner optically 
thick disk ($ 10 r_s < r < 30 r_s$). The active regions in the hot corona models
may then be uniformly situated over the inner region of an
optically thick disk ($r < 20 r_s$) where most of the gravitational energy
is dissipated.
However, the presence of a Comptonizing medium would raise questions
regarding it's dynamical support, stability, geometry and origin.   
  
\section{The Cloud Model}

We consider a uniform cloud of radius $R$ surrounding the
black hole with scattering optical depth, $\tau_{sc} \approx 5$. For
no absorption edges to be observed the gas has to be highly ionized. Following
Fabian \etal 1995 we first consider the outer layers of the optically thick
cloud. The outer layer here is defined such that $\tau = n_e \sigma_T \Delta R
\approx 1$, where $n_e$ is the electron density of the cloud and $\Delta R$ is the
thickness of the outer layer. Kallman \& McCray (1982) have
calculated the ionization equilibrium for such clouds provided $\tau_{sc} < 0.3$.
Since in this layer the scattering optical
depth is less than unity, their calculations are approximately valid.
Furthermore since multiple scattering is rare in this region, their calculation
for the equilibrium temperature would also be approximately correct. Their
results indicate that the outer layer of the cloud is highly ionized if 
the ionization parameter $\Xi \equiv L/n_eR^2 > 10^4$, where $L \approx 10^{43}$
ergs sec$^{-1}$ is the luminosity of the source. Thus the maximum
radius for this cloud is $R < 10^{14}$ cms ( Fabian \etal 1995).
Iwasawa \etal 1996 report that the line is variable on a time
scale less than $\onlyten{4}\sec$, which also implies that the size
of the cloud is $ R < 3 \times 10^{14}$ cms.
For a $\onlyten{7}\msol$ black hole, this would correspond to $\approx 30 r_s$.
We note that this upper limit on the radius (relative to $r_s$) of the Comptonizing cloud
depends on the assumed mass of the black hole.
For example, if the black hole mass is $\onlyten{6}\msol$, the limit
would be $\approx 300 r_s$. 
For these conditions,  Kallman \& McCray (1982)
calculate the equilibrium temperature to be $\approx 1$ keV. This temperature
is higher than the temperature implied from the line profile (Fabian \etal 
1995). However, this
is the outer layer temperature. The temperature of the inner regions of
the cloud (where most of the Comptonization occurs) is expected to be
less than this value, due to multiple scattering ( see below).

Estimating the ionization state of the gas in the inner region of
the cloud is complicated, due to multiple scatterings of the photons.
The calculations of Kallman \& McCray (1982) are no longer valid.
Since this region is closer to the X-ray source the photon flux is
higher here; moreover the photon density inside is also expected to be $\tau$
times higher than the value calculated using the free streaming 
approximation. From these arguments it is expected that the cloud
should be more ionized in the interior than the outer layer. On the
other hand the equilibrium temperature in the inner region is expected 
to be smaller due to multiple scatterings. 

From the above arguments, Fabian \etal (1995)
concluded that
a cloud with radius $R > 10^{14}$ cms, will not be
highly ionized. This upper limit
on the size
is compatible with the observed variability of the source. Thus if
a Comptonizing cloud surrounds this source its radius should be less 
than $10^{14}$ cms.

\section{The effects of Comptonization}

In this
paper we consider the effect of a uniform Comptonizing cloud with $R = 10^{14}$
cms on the X-ray
continuum and line shape. As an extreme case, we neglect gravitational
and Doppler broadening, which would be justified if the black hole
mass is $\approx \onlyten{6}\msol$.
The  spectrum of the central source before the Comptonization is assumed to be an X-ray
power-law with an exponential cutoff, a UV bump and a reflected component.
The UV photons are most probably the intrinsic emission of the 
reflecting surface. Hence, for consistency, the UV photon source is also
assumed to be within the Comptonizing cloud. 
The output spectrum after the Comptonization is
obtained by solving the Kompane'ets equation with this  spectrum
as the source. We take into account bremsstrahlung photon production  and absorption 
in the cloud. However, we find that neither of these is
important for the parameters chosen here. The distance to the source is taken
to be $50\mpc$  (red-shift $z = 0.008$ with $\hubble{50}$).

The temperature of the cloud was calculated by balancing the input
radiative power to the output power. For this purpose, the Kompane'ets
equation was solved by assuming some value for the temperature. Then the luminosity in the
output (\ie Comptonized) spectrum was compared to the intrinsic luminosity of
the central source. The temperature was then varied till the luminosities
were equal. This equilibrium temperature is a function of the scattering
optical depth $\tau$ in the cloud and the estimated UV flux of the source.

In Figure 1, the dashed  line shows the intrinsic spectrum assumed by us. 
The dotted line is the best fit X-ray spectrum  of MCG-6-30-15 as observed by Ginga 
(Nandra \& pounds 1994, Table 8, Set 3),
in the $2-18\kev$ range. Based on this we take the power-law photon index of 
the spectrum to be 2.07.  We assume that the spectrum is exponentially cutoff 
at $E_c = 600\kev$ and that the reflection parameter $R_{\rm ref} \equiv \Omega/2\pi = 1.2$.  
The UV bump is assumed to have a nearly black-body like shape,
\begin{equation}
F_{uv} = A E^{0.3} exp( -E/E_T )\unit{photons/sec/cm^2/keV}.
\end{equation}
Here $E$ is the photon energy and $E_T$ is assumed to be $20\ev$. 
The form is chosen to match with the  average UV spectral slope of 1.3
observed in Seyferts (Kinney \etal 1991). The UV flux at
$1550\AA $, \ie $8\ev$ is observed
to be correlated to the X-ray flux at $2\kev$ ( Gondhalekar, Rouillon-Foley \&
Kellett 1996). We have estimated the UV flux using the analytical expression
fitted to this correlation. The dot in Figure 1 represents this UV flux 
($\approx 5 \times 10^{-26}$ ergs s$^{-1}$ cm$^{-2}$ Hz$^{-1}$) for
the X-ray flux ($\approx 5 \times 10^{-29}$ ergs s$^{-1}$ cm$^{-2}$ Hz$^{-1}$)  observed by Ginga. 

 The temperatures obtained
for different optical depths, obtained using  the input spectrum assumed
for this source, are shown in Table 1. The temperature also
depends on the estimated UV flux of the source.

\btable
\caption{Temperatures for different values of optical depth.}
\begin{tabular}{|c|c|}          \hline    \hline
Optical Depth     & Temperature    \\ \hline
5                 & 0.18            \\
3                 & 0.22           \\
1                 & 0.23           \\ \hline
\end{tabular}
\etable 

The solid line in Figure 1 is the Comptonized spectrum from a cloud
with $\tau = 3$ and the calculated  temperature of $0.22\kev$. As shown
below the system seems to be in this state for most of its time. In the
energy band $2 - 18\kev$ the spectrum is nearly equal to the intrinsic
spectrum. Thus in the Ginga band the cloud does not produce any
observable signatures. At $100\kev$ the flux is reduced by about
$60\%$. The averaged Seyfert 1 spectrum from OSSE has been 
presented by Gondek \etal (1996). The flux times energy  at 100 keV lies
between 0.02 and $0.03\unit{\kev\,s^{-1}\cm^{-2}}$, which is just compatible
with the predicted value here. However, comparing with an averaged
spectrum may be misleading since not all the Seyferts used in the
sample have a broad Iron line and hence in the framework of this
model do not have optically thick clouds. Moreover,
the OSSE and ASCA observations are not simultaneous. The
line width is variable which in this framework implies that 
the optical depth of the cloud also varies. Thus the Comptonized X-ray
spectrum of MCG-6-30-15 is not in conflict with the present observations.

However, simultaneous broad band (2 - 200 keV) observations of this
source would be able to test the cloud model.
In particular, a break in the spectrum ( see figure 1)
around 50 keV would indicate the presence of a Comptonizing cloud.
On the other hand if such a break is not observed and 
{\it simultaneous } data shows that the line is very broad, that
would argue against a Comptonizing cloud. Thus simultaneous
observations of this source by ASCA and RXTE or BeppoSAX may
be able to test the Comptonizing cloud model. It should be noted
that a significant break at 30 keV, has not been observed in any Seyferts
so far (Gondek \etal 1996).
 
We assume that the intrinsic Iron line is a Gaussian with dispersion
$\sigma_I = 0.2\kev$. In figure 2 (a) we show this intrinsic line
(dotted) and the Comptonized line (solid). Here $\tau = 5$ and
the temperature is $ 0.18\kev$. The dashed line is the best fit
double Gaussian model to the data obtained by Ginga for one set
of observations by Iwasawa \etal (1996). This was named as the
deep minimum phase and showed the maximum broadening. Although
the Comptonized spectrum does not compare well with the best
fit spectrum, some of the salient features 
are present: namely, there is an extended red wing and a sharp
drop towards the blue side. In fact, comparison of the Comptonized spectrum
with the unfolded photon spectrum shown in  Figure 7 of Iwasawa \etal (1996),
rather than with their best fit, reduces the discrepancy.
In Figure 2 (b), $\tau = 3$ and the
temperature is $0.22\kev$. The Comptonized spectrum is compared with
the best fit double Gaussian fit obtained for the Intermediate flux
state (Iwasawa et al 1996). The system was in this state for five
of the seven sets of observations. Figure 2 (c) shows the Comptonized
spectrum when $\tau = 1$ and temperature is $0.23\kev$ compared
with the bright flare phase.

We note that the broadened spectrum due to Comptonization is not similar
to the double peaked spectrum expected from  inner disk emission. Higher
resolution observations of the line would be able to differentiate between 
the two scenarios. In particular, if the double peaked nature of the line
is observed (not just the broadening), then the emission would indeed be
produced close to a spinning black hole. On the other hand, a single
broadened line would be consistent with emission from a 
Comptonizing medium. 

Spectral fitting of the X-ray band for MCG-6-30-15 requires the
presence of a  variable warm absorber with hydrogen column density
$N_H\sim\ten{6}{22}\cm^{-2}$. In the simplistic model
presented here we have considered a uniform cloud with a definite
radius. In reality, the density of the cloud  probably decreases
with radius and the low density region of the cloud extends to greater
radii than $10^{14}$ cms considered here. We speculate that the warm
absorber could have its origin in this extended region of the cloud.

\section{Discussion}

In this paper, we show that an optically thick
Comptonizing cloud with radius $R \approx 
10^{14}$ cms, could contribute significantly to the broadening of
an intrinsic iron line. The gas would probably be highly ionized
and hence no absorption edges should be observed (Fabian \etal 1995). 
We find that taking into account the UV radiation from the
source and multiple scattering in the medium, the equilibrium temperature of the cloud is $kT \approx 0.2$ keV. The shape of the Comptonized line
from such a medium is similar to the one observed.
If the black hole mass is $\approx \onlyten{6}\msol$
the Compton broadening could be dominant. For a $\onlyten{7}\msol$ black 
hole, gravitational/Doppler effects would also contribute.

The present
hard X-rays observations are compatible with the spectrum expected
from such a cloud. However, simultaneous broad band observation
of this source (e.g. with ASCA, RXTE and BeppoSAX ) 
will be able to test the presence of such a Comptonizing cloud.

We emphasis that the spectral shape of the Comptonized line is
different from the disk emission non-Comptonized one. In particular,
the disk emission line is double peaked, while the Comptonized
spectrum is continuous. If the line is observed to be double peaked,
that would indicate that Comptonization is not dominant. However,
direct evidence for the presence of a Comptonizing
medium would be obtained, if the line is observed to be continuous.
In such a scenario, the cold accretion disk need not extend to
the last stable orbit, the black hole may not be spinning and 
the hard X-ray source need not be situated very close to the black hole.
Thus, if there is a Comptonizing cloud, the X-ray producing
regions could still be described by either an hot inner disk
or a coronae (active region) on top of a cold disk.
It is also possible that both Comptonization and gravitational red-shift
contribute significantly to the broadening of the line.

\acknowledgments

The authors would like to thank A. Fabian for useful criticism
of an earlier draft and A. A. Zdziarski for comments and discussion.

\clearpage

\figcaption{ The intrinsic spectrum of the central source (dashed
line) and the Comptonized spectrum from a cloud with $\tau = 3$ 
(solid line). The spectral fit to Ginga data in the energy band
$2-18$ keV (dotted line) and the estimated UV flux (dot).\label{fig1}}

\figcaption{ The intrinsic narrow Iron line from the central source (dotted
line) and the Comptonized Iron line from a cloud (solid line). The
best fit double Gaussian model fit to data (dashed line). (a) for
the deep minimum phase compared to line from a cloud with $\tau = 5.0$.
(b) for the intermediate flux  phase compared to line from a cloud 
with $\tau = 3.0$. (c) for the high flare  phase compared to line 
from a cloud with $\tau = 1.0$.
\label{fig2}}

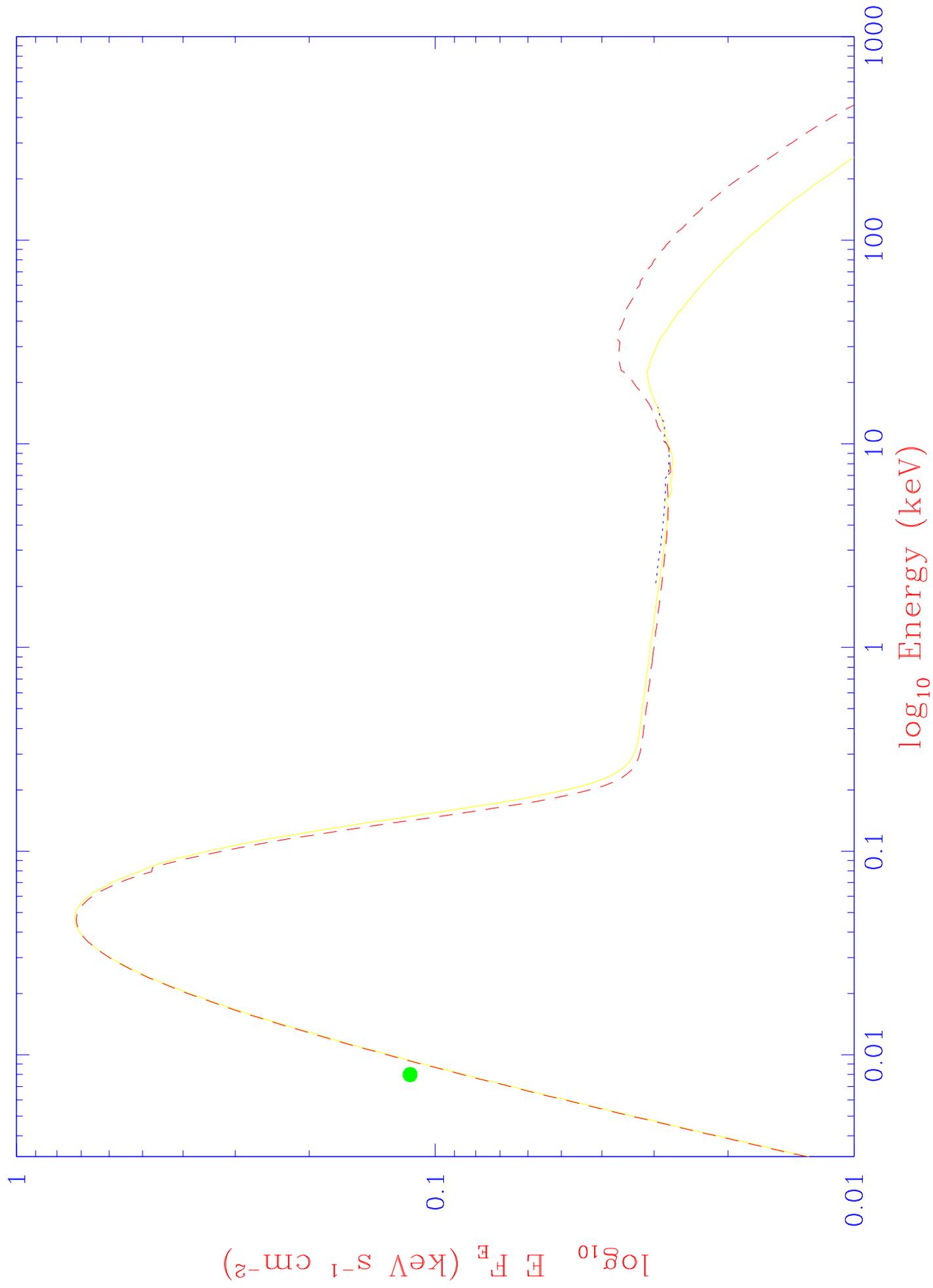
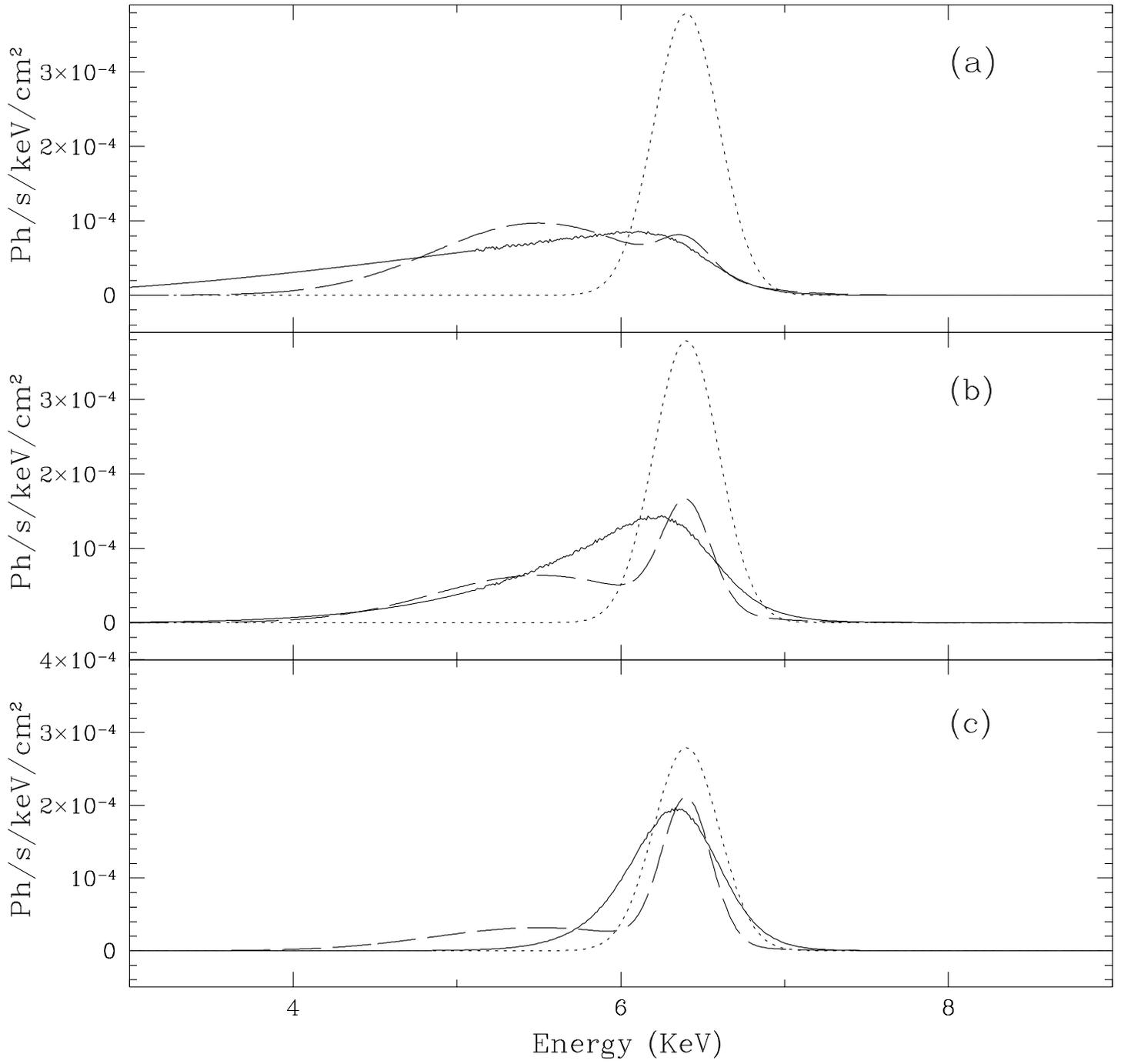
\end{document}